\documentclass[twocolumn]{emulateapj}

\usepackage[colorlinks=true,linkcolor=blue,citecolor=blue]{hyperref}
\usepackage{amsmath}
\usepackage{bbding}
\usepackage[dvips]{color}
\usepackage{graphicx}

\newcommand{\rprs}{R_\mathrm{p}/R_*}

\defcitealias{yelle}{Y04}

\shorttitle{CoRoT-1b Transmission Spectrum}
\shortauthors{Schlawin, Zhao, Teske, Herter}

\begin{document}


\title{A 0.8-2.4 Micron Transmission Spectrum of the Hot Jupiter CoRoT-1b}


\author{E. Schlawin\altaffilmark{1}, M. Zhao\altaffilmark{2,3}, J. K. Teske\altaffilmark{4}, T. Herter\altaffilmark{1}}
\altaffiltext{1}{Astronomy Department, Cornell University, Ithaca NY 14853}
\altaffiltext{2}{Department of Astronomy, Pennsylvania State University, University Park PA 16802}
\altaffiltext{3}{Center for Exoplanets and Habitable Worlds, University Park PA 16802}
\altaffiltext{4}{Astronomy Department, The University of Arizona, Tucson AZ 85721}


\bibliographystyle{apj}

\begin{abstract}
Hot Jupiters with brightness temperatures $\gtrsim$2000K can have TiO and VO molecules as gaseous species in their atmospheres. The TiO and VO molecules can potentially induce temperature inversions in hot Jupiter atmospheres and also have an observable signature of large optical to infrared transit depth ratios. Previous transmission spectra of very hot Jupiters have shown a lack of TiO and VO, but only in planets that also appear to lack temperature inversions. We measure the transmission spectrum of CoRoT-1b, a hot Jupiter that was predicted to have a temperature inversion potentially due to significant TiO and VO in its atmosphere. We employ the multi-object spectroscopy (MOS) method using the SpeX and MORIS instruments on the Infrared Telescope Facility (IRTF) and the Gaussian Process method to model red noise. By using a simultaneous reference star on the slit for calibration and a wide slit to minimize slit losses, we achieve transit depth precision of 0.03\% to 0.09\%, comparable to the atmospheric scale height but detect no statistically significant molecular features. We combine our IRTF data with optical CoRoT transmission measurements to search for differences in the optical and near infrared absorption that would arise from TiO/VO. Our IRTF spectrum and the CoRoT photometry disfavor a TiO/VO-rich spectrum for CoRoT-1b, suggesting that the atmosphere has another absorber that could create a temperature inversion or that the blackbody-like emission from the planet is due to a spectroscopically flat cloud, dust or haze layer that smoothes out molecular features in both CoRoT-1b's emission and transmission spectra. This system represents the faintest planet hosting star ($K$=12.2) with a measured planetary transmission spectrum.
\end{abstract}


\keywords{radiative transfer, planets and satellites: individual (CoRoT-1b), stars: individual (CoRoT-1), (stars:) planetary systems}

\section{Introduction}\label{intro}

Transiting hot Jupiters are among the most observationally favorable sources for measuring atmospheric composition, global winds, temperature inversions and disequilibrium chemistry \citep[e.g.,][]{pont13,snellen10,rogers09,moses11}. Their large physical radii, frequent transits, high temperatures and large radial velocity amplitudes permit both the measurement of physical parameters (mass, radius, orbital elements) and the ability to test atmospheric models. The primary transit, when the planet goes in front of its host star, and the secondary eclipse, when the planet goes behind, are valuable opportunities to spectroscopically characterize the atmosphere. These spectra can be compared with models to determine mixing ratios of atmospheric gases, clouds, scatterers and/or aerosols. Furthermore, high quality spectra can be used to constrain the formation of exoplanets \citep[e.g.,][]{spiegel12}, the extent of equilibrium/disequilibrium chemistry \citep[e.g.,][]{moses11}, vertical mixing \citep[e.g.,][]{visscher11} and put the Solar System in context.

Transmission spectra and emission spectra of hot Jupiter atmospheres have already been used to detect Na \citep{charbNa}, K \citep{sing11}, Ca \citep{astudillo13}, H \citep{viddisc}, H$_2$O \citep[e.g.,][]{deming13,birkby13}, CO \citep[e.g.,][]{snellen10} and possibly CH$_4$, \citep[though see \citet{gibson12}]{swain08}. Furthermore, emission and transmission spectra have been used to constrain the mixing ratios of these atoms and molecules. Of considerable interest is the relative abundances such as the C/O ratio \citep{teske13,madhusudhan12,madhusudhan11}, which gives clues as to the formation of planets such as circumstellar disk composition and location within the disk \citep[e.g.,][]{oberg11,moses13}.

Infrared observations of prominent molecular bands in hot Jupiters during secondary eclipse are used to infer an atmospheric temperature profile \citep[e.g.,][]{Line2013}. The level of emission by gases of upper layers as compared to lower levels indicates their relative temperatures. For example, the brightness temperature of the 4.5 $\mu$m {\it Spitzer} band is expected to be higher than the 3.6 $\mu$m band for temperature-inverted planets because it encompasses several molecular bands that are high in opacity (and high in altitude), whereas the 3.6 $\mu$m band sees deeper in the atmosphere \citep{knutson2010correlation}.

Broadly, hot Jupiter atmospheres have been classified into (1) planets that have temperatures that decrease with altitude for observable pressures and (2) planets that contain a temperature inversion or stratosphere at observable pressures. We include an isothermal (constant temperature with altitude) in the later case. One possible explanation for the bifurcation into theses profiles is that TiO and VO absorption of stellar flux creates temperature inversions in some planets and not others \citep{hubeny03,fortney08twoclass}. An alternative explanation is that the observational techniques to infer temperature inversions (like the 4.5 $\mu$m to 3.6 $\mu$m brightness ratio) are actually sensing the difference between clear atmospheres and dusty atmospheres, such as has been observed in HD 189733b \citep{pont13,evans2013}. Recently, spectro-photometry of HAT-P-32b \citep{gibson13clouds}, HAT-P-12b \citep{line13hatp12}, WASP-17b \citep{mandell2013wfc3}, GJ 1214b \citep{kreidberg14}, GJ 436b \citep{knutson2014gj436} and phase curves of Kepler-7b \citep{demoryClouds} indicate that clouds and hazes may be common in exoplanet atmospheres.

The very short period hot Jupiters, such as WASP-12b \citep[$P$=1.09 days]{hebb09}, WASP-19b \citep[$P$=0.79 days]{hebb10}, HAT-P-32b \citep[$P$=2.2]{hartman2011} and CoRoT-1b \citep[$P$=1.51 days]{barge08}, are in the temperature regime where TiO and VO may be abundant atmospheric constituents \citep{fortney10} -- their brightness temperatures are respectively 3600 K \citep{crossfield2012}, 2700 K \citep{abe2013}, and 2500 K \citep{deming11}. TiO and VO are molecules that are so sensitive to the C/O ratio that their abundances decreases by a factor of $\sim$100 going from C/O=0.54 (solar) to C/O=1 \citep{madhusudhan11ApJ}. Their presence should be accompanied by a greater radius for the optical wavelengths ($\sim$450 to $\sim$850nm) than for infrared wavelengths ($\gtrsim$1000nm) \citep{fortney10} and could explain the bifurcation scheme of planets into temperature inverted and non-temperature inverted planetary atmospheres \citep{hubeny03,fortney08twoclass}.

Recent transmission spectroscopy observations have measured the level of TiO and VO in the atmospheres of WASP-19b's, WASP-12b and HAT-P-32b. The first two planets lack temperature inversions \citep{Line2013}, so TiO and VO should be removed from their higher altitudes. Indeed, \citet{huitson13} found that the transmission spectrum of the hot Jupiter WASP-19b has low levels of TiO as compared to theoretical models with solar abundances and local chemical equilibrium. \citet{mancini13} also find that WASP-19b's transmission spectrum is consistent with models without TiO/VO absorption. Observations of WASP-12b during primary transit and secondary eclipse were consistent with TiO/VO and TiH absorption \citep{swain2013,stevenson2013wasp12} but including aerosols in the calculated transmission models and adding HST optical data suggest that TiO/VO are not dominant absorbers \citep{sing2013}. It is possible that WASP-12b's TiO and VO are trapped on the planet's nightside \citep{sing2013}.
HAT-P-32b's transmission spectrum also shows a lack of strong TiO/VO features, possibly due to gray-absorbing clouds \citep{gibson13clouds}.

The hot Jupiter CoRoT-1b, orbiting a 5960 K effective temperature $V$=13.6 star \citep{barge08}, is better matched with models that include a temperature inversion \citep{rogers09,gillon09,zhao12} or an isothermal profile \citep{deming11}. It is thus is a potential candidate for strong observable signatures of TiO/VO. This makes CoRoT-1b a useful comparison planet to WASP-19b and WASP-12b because it has a similarly high brightness temperature ($>$2000K) but a different temperature profile. \citet{deming11} additionally find that CoRoT-1b's secondary eclipse spectrum is well fit by a blackbody, which could indicate an isothermal temperature gradient or, alternatively, a high altitude dust such as been found in HD 189733b \citep{pont13}.

CoRoT-1b is favorable for characterization due to its large radius \citep[$R_\mathrm{p}$ = 1.49 $R_\mathrm{Jup}$][]{barge08}, high temperature \citep[$T_\mathrm{blackbody}$ = 2450 K][]{deming11} and moderate mass $M_\mathrm{p}$ = 1.03 $M_\mathrm{Jup}$, which combine to give it a large scale height $H=kT/\mu mg \approx 0.01 R_\mathrm{Jup}$ where $k$ is Boltzmann's constant, $T$ is the kinetic temperature, $\mu$ is the mean molecular weight =2.3 for a solar mixture, $m$ is one atomic mass unit and $g$ is the local gravitational acceleration. Furthermore, there is a nearby reference star close in brightness and color (within 0.7 magnitudes in the $J$, $H$ and $K$ bands) that permits characterization with the multi-object spectroscopy (MOS) method \citep{bean10,sing12,gibson13,bean13}.

The MOS method is to divide a target star spectrum by one (or an average of several) reference stars to correct for variability in telluric (Earth's) transmission and the response of the instrument. Close proximity of a reference star to the target provides an advantage for calibration, as their atmospheric turbulence and telluric fluctuations are highly correlated. The reference stars' spectra are obtained simultaneously either with multiple slits or, as in our observations, a long slit that includes both the planet hosting star and the reference star.

One observational challenge with the CoRoT-1 system is its faintness at $K$=12.2. This makes it difficult to obtain sufficient signal to noise for high resolution measurements but we demonstrate that the Infrared Telescope Facility (IRTF) with SpeX and MORIS instruments in a low resolution prism mode (with no diffraction grating) can achieve high precision characterization down to this faint magnitude. We present a 0.8 $\mu$m to 2.4 $\mu$m transmission spectrum to constrain the presence of infrared absorbing molecules and measure the optical/near IR radius slope as compared to TiO/VO absorption.

\begin{figure}[!ht]
\begin{center}	
\includegraphics[width=0.5 \textwidth]{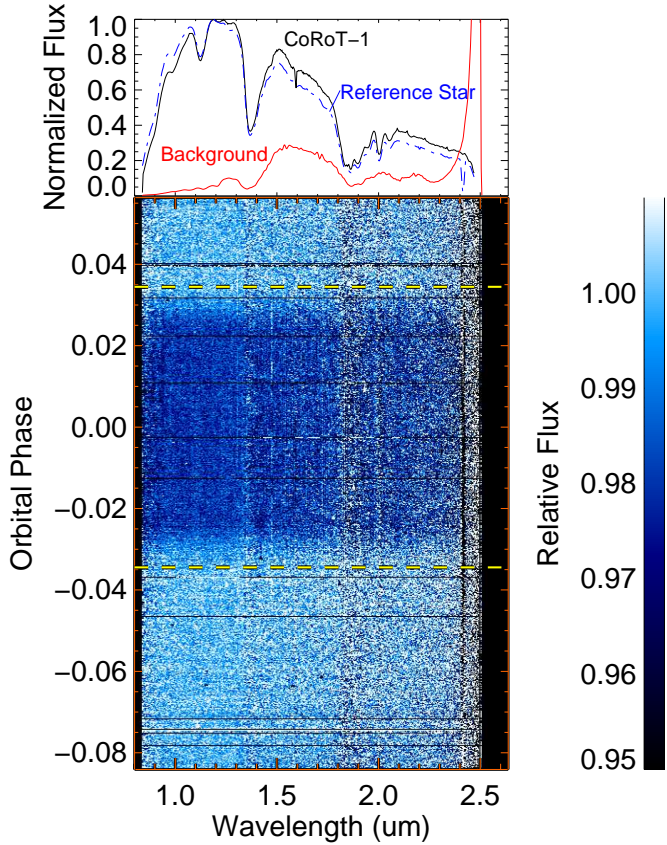}
\end{center}
\caption{{\it Top} Normalized spectra for the planet hosting star, the reference star and background for Jan 04, 2012 indicate the regions where there are strong telluric absorption features, strong background emission and detector effects (spurious absorption features at 2.41 $\mu$m and 1.58 $\mu$m). {\it Bottom}: Dynamic spectrum for the night of Jan 04, 2012. Each row in the image is a single spectrum of CoRoT-1 divided by the reference star and re-normalized with a linear baseline. The transit (encompassed by horizontal yellow dashed lines at ingress/egress) is clearly detected in all wavelength channels save the ends of the spectrograph.}\label{dynamicS}
\end{figure}

\section{Observations}\label{mcond}

We observed CoRoT-1b with the SpeX instrument \citep{rayner03} on the Infrared Space Telescope Facility in a low resolution prism mode. When the large 3'' x 60'' slit is placed on CoRoT-1, the actual resolution for the target star is set by the point spread function at $R\approx$80. A reference star -- 2MASS 06482020-0306339 -- was placed simultaneously on the slit to correct for telluric transmission variations as well as correlated (common mode) instrumental variations. The 3'' x 60'' slit was selected to minimize slit losses but it still serves to reduce the background levels as compared to a completely slit-less instrument. The reference star with $J$=11.72, $H$=11.54, $K$=11.50 is slightly brighter than CoRoT-1 $J$=12.46, $H$=12.22, $K$=12.15 as determined from 2MASS \citep{skrutskie06} so that the photon noise of the planet host star dominates the photon noise of the measurement. We kept the exposure times short to keep the counts of the two objects well within the linear regime of the detector. At the same time, their fluxes are close enough so that flux-dependent non-linearity is negligible.

We observed CoRoT-1 for 3 nights on the UT dates of Dec 23, 2011 (full transit), Dec 29, 2011 (half transit) and Jan 04, 2012 (full transit). The first half of Dec 29, 2011 was lost due to high wind ($>$45 MPH) and closure of the telescope. The remainder of the Dec 29, 2011 night was affected by large seeing fluctuations from 0.9'' to 1.5''. For the full transits, the 2.5 hour transit duration was straddled by 30 to 120 minutes of out-of-transit observations to establish a baseline flux level. Table \ref{obsSummary} lists the exposure times and number of exposures obtained for the three transits.

We also used MORIS, a high-speed, high-efficiency optical camera \citep{Gulbis2011} simultaneously with SpeX to obtain photometry at the Sloan $z'$ band for CoRoT-1. We used a 0.9 $\mu$m dichroic to split visible light short-ward of 0.9 $\mu$m into the MORIS beam path.  
The field of view of MORIS is similar to the guide camera of SpeX (1' x 1' arcmin), permitting us to include two reference stars in addition to CoRot-1 on the MORIS detector. We used short exposures of 5s and 10s to ensure the fluxes were well within the linear regime of the camera. The observing log of MORIS is also included in Table \ref{obsSummary}. Photometric data reduction was carried out following the pipeline and steps of \citet{zhao12}. The total flux of the two reference stars (2MASS 06482101-0306103 and 2MASS 06482020-0306339) was used for flux calibration. We determined that an aperture size of 36 pixels (corresponding to 4.1'' for a pixel scale of 0.114"/pixel) and a 35-pixel wide background annulus provided the best light curve precision for all 3 nights, although aperture sizes with $\pm$5 pixels gave essentially the same results. 

The spectral images were reduced with standard \texttt{IRAF ccdproc} procedures with four to eight flat frames, dark subtraction from identical exposure time frames and one to two wavelength calibration frames. Wavelength calibrations were performed with a narrower (0.3''x60'') slit to better centralize the Argon emission lines. Additionally, we rectify all science images using the Argon lamp spectrum as a guide to make sure all vertical columns in the image correspond to individual wavelengths.

Simultaneous $H$+$K$ band exposures were made with the infrared guider on SpeX to ensure good alignment of the target and reference star. The stars are visible as reflections off the slit, permitting a simultaneous check that the stars are centered during spectrograph science exposures. In addition to the reflections from the slit, nearby additional reference stars off the slit were also evaluated for centroid motions. The centroid motions show that guiding using the $H$+$K$ guider was accurate to within 0.3'', minimizing slit loss errors in the spectrograph. No correlations are visible between telescope shifts (measured from $H$+$K$ images) and the individual target and reference stars' fluxes or ratio spectrum between the planet host and reference star.

We extracted all of the spectra with the \texttt{twodspec} procedures in \texttt{IRAF} \citep{tody93,tody86}. We used a centered aperture of 15 pixels (2.3'') with optimal extraction (spatial pixels weighted by S/N ratio) on the planet host star and reference star (FWHM $\approx$5 to 9 px or 0.8'' to 1.4'') with a third order Legendre polynomial background subtraction from 89 pixels on each side of the spectrum. These extraction and background sizes were chosen experimentally so as to produce the smallest standard deviation of out-of-transit flux in the final time series.  The fact that the highest precision was obtained with a 2.3'' aperture size shows that the 3'' slit width is sufficiently wide to make slit losses negligible.

For each exposure, the CoRoT-1 system's spectrum was divided by the reference star to correct for variable transmission and response of the instrument. This is the same long slit/multi-object method applied by \citet{sing12}, \citet{bean10}, \citet{bean13} and \citet{gibson13}. Figure \ref{dynamicS} shows a dynamic spectrum from the night of Jan 04, 2012 using the reference star division and then re-normalizing each time series by a linear out-of-transit baseline. The linear baseline division is only used for illustrative purposes in this figure and not the parameter extraction described in Section \ref{paramExt}. Each of the 475 wavelength channels clearly shows the transit except for the ends of the spectrograph due to low response and high thermal background at the larger wavelength end. The telluric transmission above IRTF at Mauna Kea is high enough that transit measurement is still possible between the $J$, $H$ and $K$ telluric windows.

\begin{table}
	\begin{center}
	\begin{tabular}{l*{5}r}
	UT Date	     & $t_\mathrm{spec}$	& $D_{spec}$	& $N_\mathrm{spec}$ & $ t_\mathrm{phot}$ & $N_\mathrm{phot}$  \\
	 		     & (s)				& 			&				&	(s)			& \\
	\hline \hline
	Dec 23, 2011 	& 10.0			& 49\%		& 813 			&  5				& 2636 \\ 
	Dec 29, 2011 	& 15.0			& 51\%		& 233 			& 10				& 691 \\ 
	Jan 04, 2012	& 15.0			& 51\%		& 600 			& 5				& 3319 \\
	\end{tabular}
	\end{center}
	\caption{\rm Summary of the 2.5 transits observed for CoRoT-1b including the exposure time for SpeX spectra $t_\mathrm{spec}$, number of spectral exposures $N_\mathrm{spec}$, spectral duty cycle $D_\mathrm{spec}$, MORIS photometric exposure time $t_\mathrm{phot}$ and number of photometric frames $N_\mathrm{phot}$. The non redundant reads were increased at longer spectrograph exposure times, thus maintaining almost the same duty cycle.}\label{obsSummary}
\end{table}

\subsection{Noise Measurements}\label{noiseSec}

\begin{figure}[!ht]
\begin{center}
\includegraphics[width=0.5\textwidth]{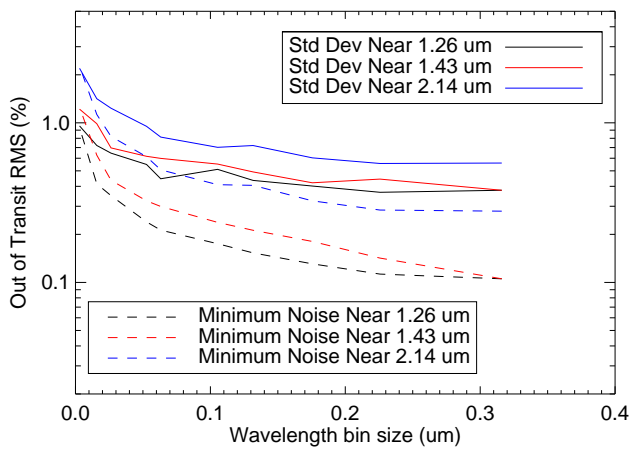}
\end{center}
\caption{Measured out of transit errors as a function of wavelength bin size for the night of Dec 23, 2011. The errors scale with minimum noise but in a non-linear way. We choose the maximum bin size possible while still resolving some broad molecular bands and use 0.17 $\mu$m bins for time series analysis. The minimum noise drops quickly for the 0.3 $\mu$m bin near 1.43 $\mu$m because there is a sharp increase in photons outside of the telluric absorption feature.}\label{wavsizeRMS}
\end{figure}

\begin{figure}[!ht]
\begin{center}
\includegraphics[width=0.5\textwidth]{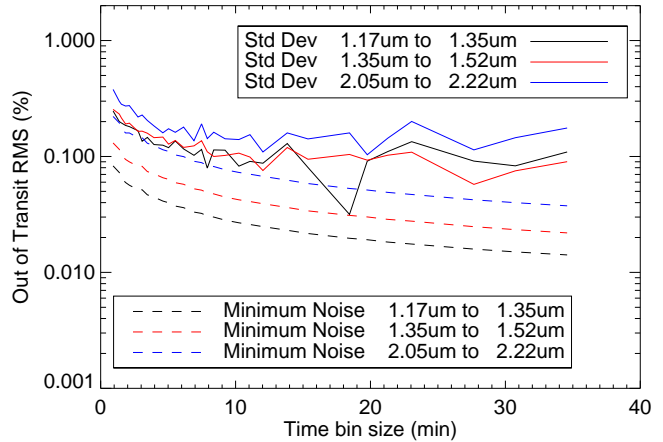}
\end{center}
\caption{Measured out of transit errors as a function of time bin size for the night of Dec 23, 2011 using 0.17$\mu$m wide bins. As with the wavelength binning, the measured noise falloff is not as sharp as with minimum noise. For 0.17 $\mu$m wide wavelength bins there is an approximate noise floor around 0.1\% and a baseline function must be used to remove long term trends. The variations in RMS for long time bins are due to small number statistics for the handful of out-of-transit flux points.}\label{timesizeRMS}
\end{figure}

The most critical part of measuring a planet's spectrum is achieving high signal to noise (S/N) ratios. Measurement errors are closely approximated by ``minimum noise'' at the highest time resolution and spectral resolution but are considerably larger when the data is binned. For this paper, ``minimum noise" includes constant read noise per pixel of the detector, shot noise of the source and shot noise of the background. Minimum noise decreases as $1/\sqrt{N}$ for N independent measurements, but we find that the measured noise falls off more slowly, as expected for high precision measurements dominated by systematics. These additional error sources are also known as time-correlated or wavelength-correlated red noise \citep[e.g.,][]{pont2006rednoise,carter2009}. Figure \ref{wavsizeRMS} and \ref{timesizeRMS} show the measured out-of-transit error as a function of bin size and also shows the minimum noise for comparison. 

For the data analysis, we used nine equally spaced wavelength bins which minimize the out-of-transit noise while still maintaining sufficient spectral resolution to resolve molecular bands. 
As expected for high precision flux measurements, the measured noise has components that do not scale as minimum noise decreases. We bin the time data slightly to $\sim$3 min long time bins for computational efficiency when doing MCMC/Gaussian Process fitting. This is still far from the noise floor seen in Figure \ref{timesizeRMS} and shorter than the planet's transit ingress duration of 22 minutes and the typical systematics $\sim$10 to $\sim$60 minutes.

\section{Light Curve Fitting}\label{paramExt}

\begin{figure}[!ht]
\begin{center}
\includegraphics[width=0.5\textwidth]{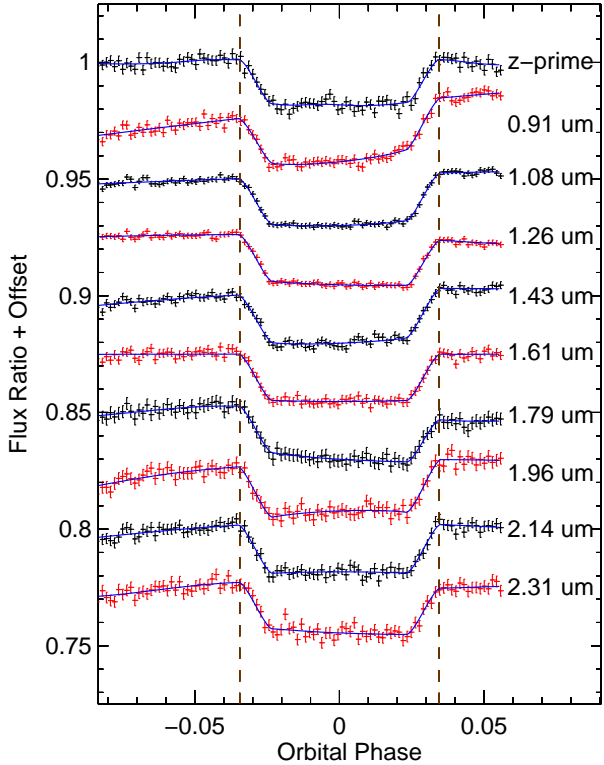}
\end{center}
\caption{Time series for the night of January 04, 2012 binned into 100 time points. The top curve (z-prime) is the MORIS photometry whereas the remaining nine are equally spaced SpeX bands. The transit light curves are fit with the \citet{magol} light curve model and with a Gaussian process error matrix \citep{gibsonGP} that includes red noise but imposes no specific baseline function for the time series.}\label{jan04timeSer}
\end{figure}

\begin{figure}[!ht]
\begin{center}
\includegraphics[width=0.5\textwidth]{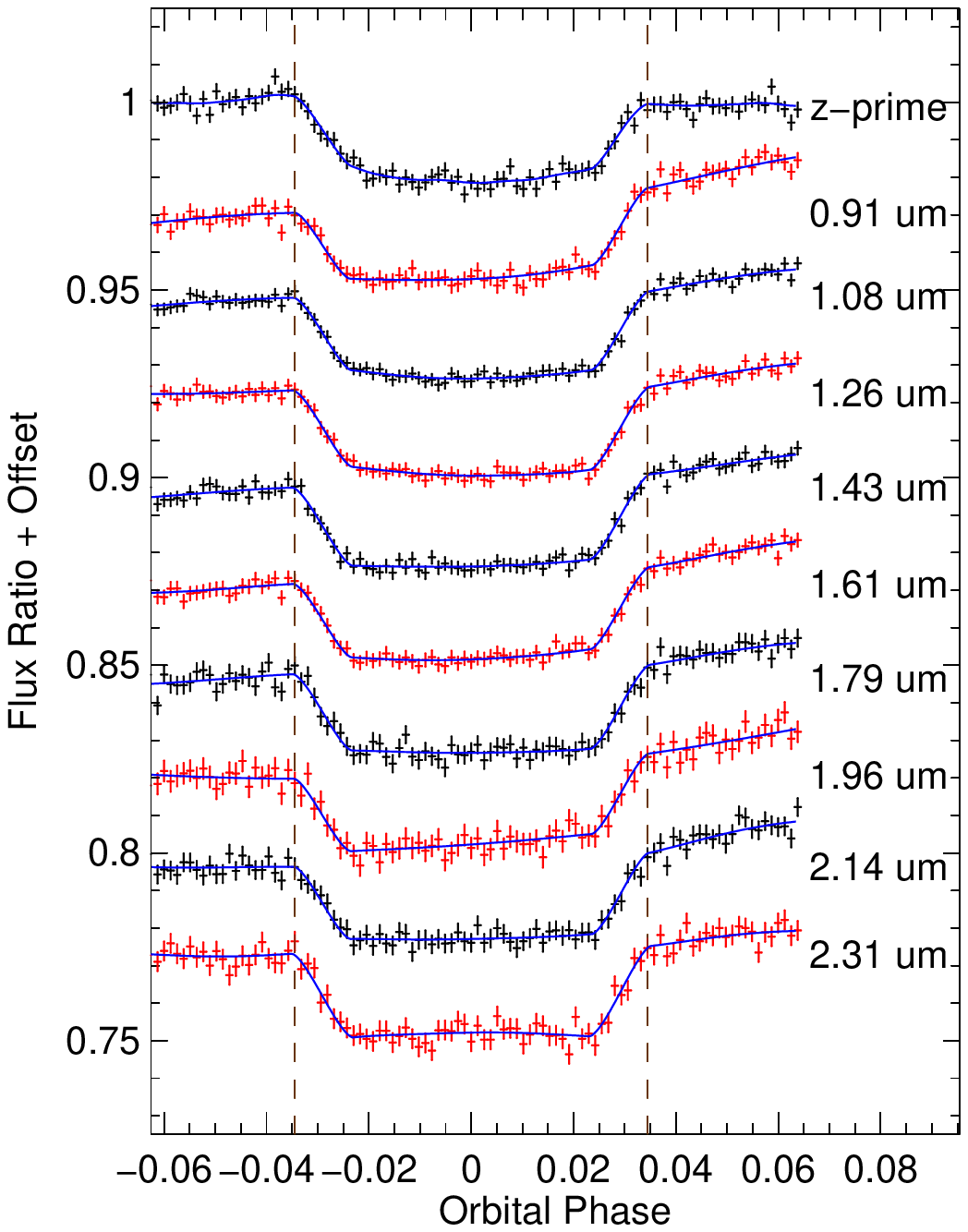}
\end{center}
\caption{Same as Figure \ref{jan04timeSer} for the night of Dec 23, 2011.}\label{dec23timeSer}
\end{figure}

\begin{figure}[!ht]
\begin{center}
\includegraphics[width=0.5\textwidth]{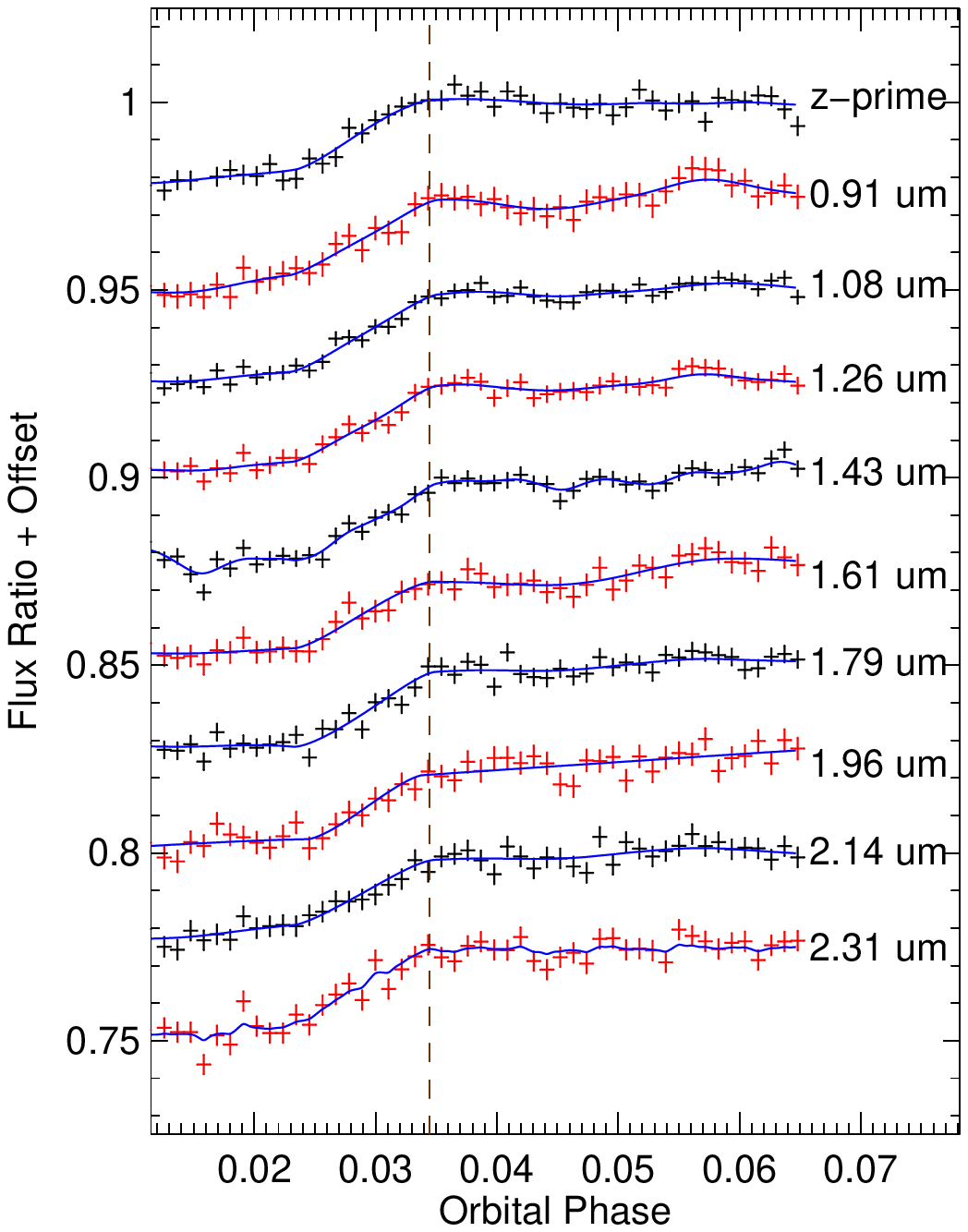}
\end{center}
\caption{Same as Figure \ref{jan04timeSer} for the night of Dec 29, 2011.}\label{dec29timeSer}
\end{figure}

As described in Section \ref{noiseSec}, all $R$=80 spectral data were binned into nine equally spaced wavelength bins and they are analyzed independently. Figure \ref{jan04timeSer} shows the time series for each wavelength bin and it can be seen that the baseline is non-linear. Figures \ref{dec23timeSer} and \ref{dec29timeSer} also show that the shape of the baseline changes from night to night. It is possible to model the baseline as a slowly varying function like a polynomial \citep[e.g.,][]{bean13} and we initially fit the light curve with a third order Legendre polynomial. The Legendre polynomials were used because their orthogonality reduces the covariance between fitted coefficients. The polynomial fits showed discrepancies between nights, so we use a non-parametric approach detailed in Section \ref{gpmod} rather than impose a specific shape on the baseline fit.

\subsection{Gaussian Process Model}\label{gpmod}
We use a Gaussian process \citep{gibsonGP,gibson13} to model red noise and the flux baseline. The advantage of the Gaussian process framework is that it does not assume that the baseline follows a pre-defined function like a polynomial where the coefficients are fitted parameters. Instead, the Gaussian process assumes the baseline and mid-transit follow a correlated normal distribution described by a covariance kernel. For repeated experiments following a Gaussian process, the actual shape of the baseline can vary from realization to realization while maintaining the same covariance kernel. The Gaussian Process method uses Bayesian model selection so that it weights against complex models to mitigate overfitting.

We use the integer form of the Mat\'{e}rn covariance kernel \citep{rasmussenGP},
\begin{multline}
C_{nm} = \Theta_0^2 \mathrm{exp}\left(-\Theta_1 \sqrt{2 \left(p+\frac{1}{2}\right)} |x_n - x_m|\right) \frac{\Gamma(p+1)}{\Gamma{(2p+1)}} \\
\times \left(\Sigma_{i=0}^p\frac{(p+i)!}{i! (p-i)!} \left( \Theta_1 |x_n - x_m| \sqrt{ 8 \left(p + \frac{1}{2} \right)} \right)^{p-i} \right)\\
+\delta_{nm} \sigma_n^2
\end{multline}\label{eq:GPkern}
where $C_{nm}$ is the covariance between data points ($x_n$,$y_n$) and ($x_m$,$y_m$), $
\Theta_0$ is a hyper-parameter describing the strength of the correlation between data points, $\Theta_1$ is the inverse time scale hyper-parameter, $p$ is the index of the Mat\'{e}rn kernel, $\delta_{nm}$ is the Kronecker delta function and $\sigma_n$ is the white-noise component of an individual point's error. This is a generalized form of the $p=1$ Mat\'{e}rn kernel used on WASP-29b transit data \citep{gibson13}. We let the $p$ parameter be another hyper-parameter with the possible values of 0, 1, 2 or infinity (a squared exponential kernel $C_{nm} = \Theta_0^2 e^{-\Theta_1 (x_n - x_m)^2/2}$) because higher values of $p$ are essentially indistinguishable from the infinity case \citep{rasmussenGP}. The four different kernels are parametrized by $\Theta_2$ with values of 0, 1, 2 and 3 for the respective values of $p$. All forms of the above kernel have correlations that decrease with separation in time. In other words, points that are close together are highly correlated but far away are less correlated. For the data series in this work, $x_n$ and $x_m$ are orbital phase and $y_n$ and $y_m$ are normalized flux. The choice of kernel does not affect the individual white noise errors which are assumed to be independent and Gaussian distributed with a standard deviation $\sigma_n$.

\begin{figure*}[!ht]
\begin{center}
\includegraphics[width=0.45\textwidth]{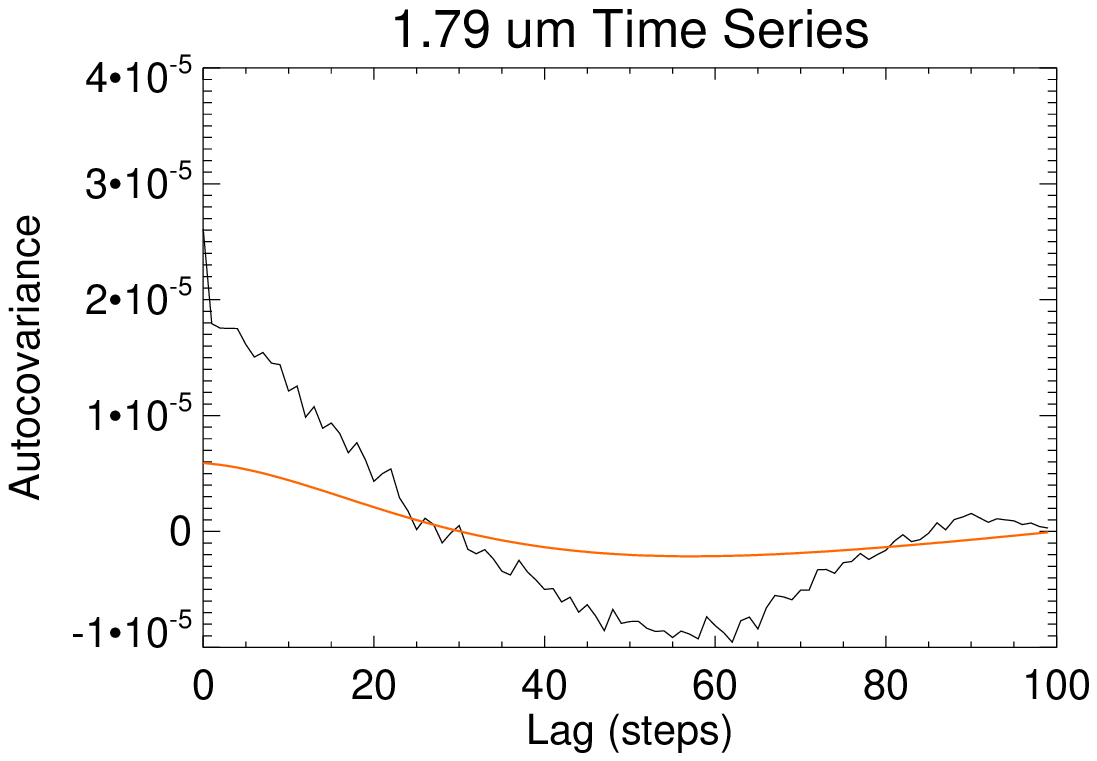}
\includegraphics[width=0.45\textwidth]{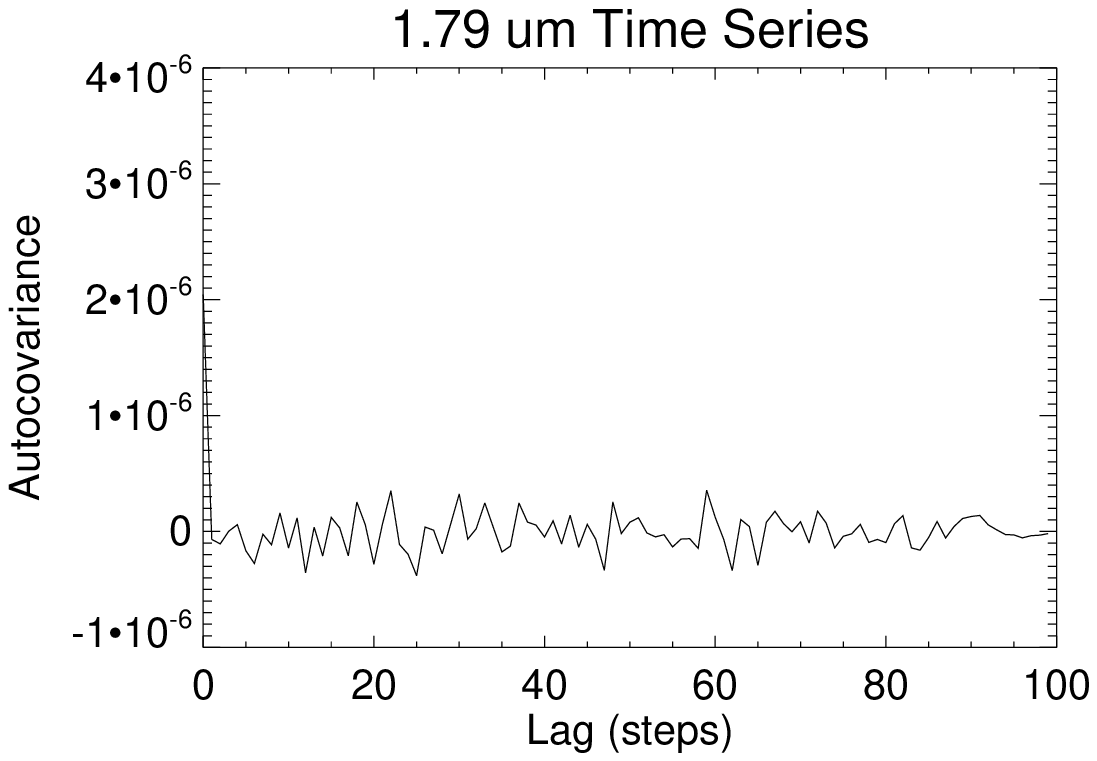}
\includegraphics[width=0.45\textwidth]{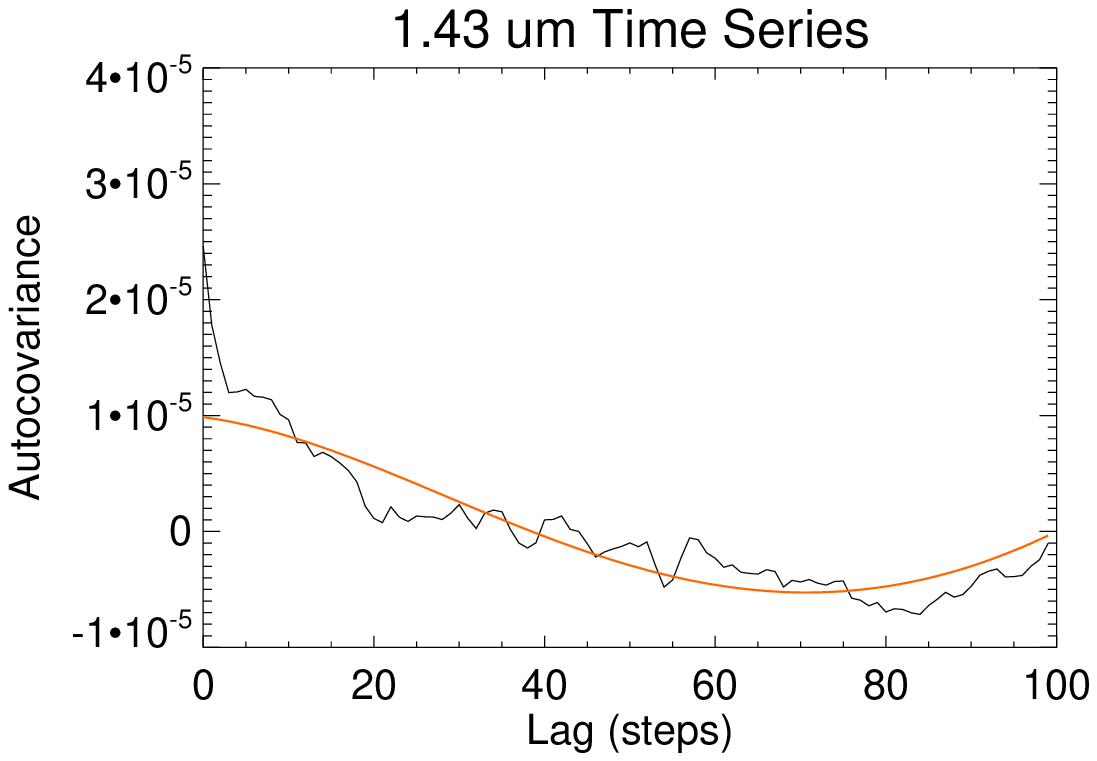}
\includegraphics[width=0.45\textwidth]{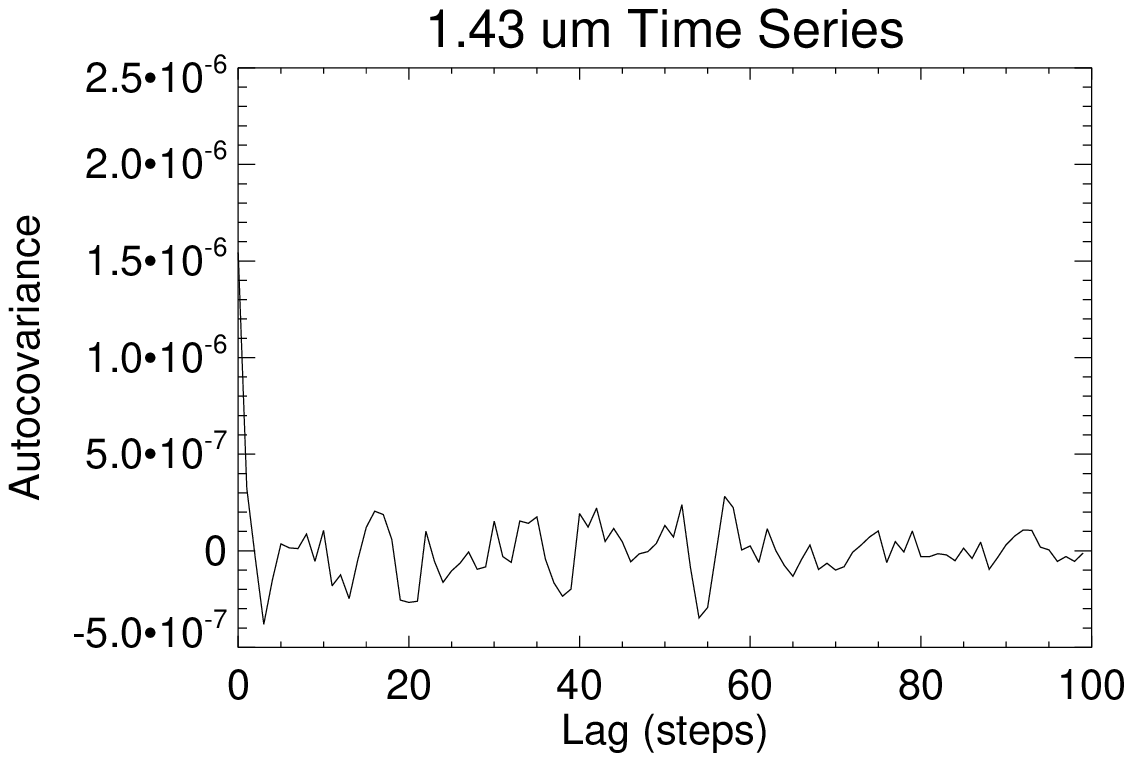}
\includegraphics[width=0.45\textwidth]{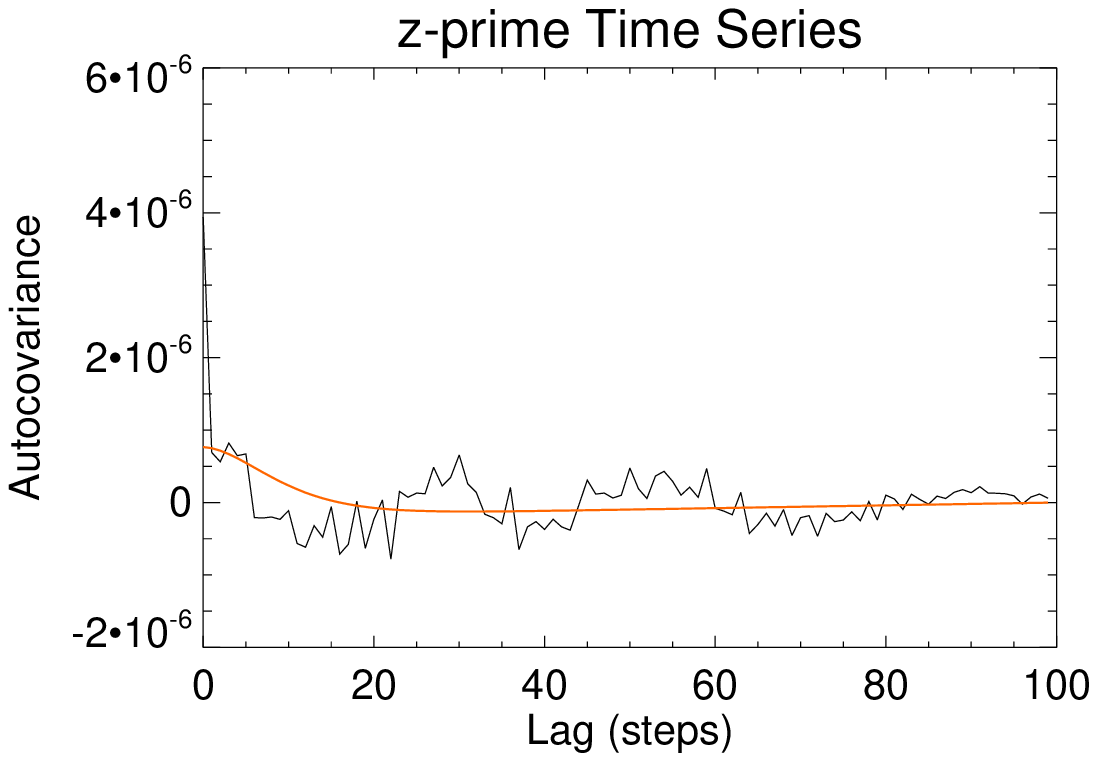}
\includegraphics[width=0.45\textwidth]{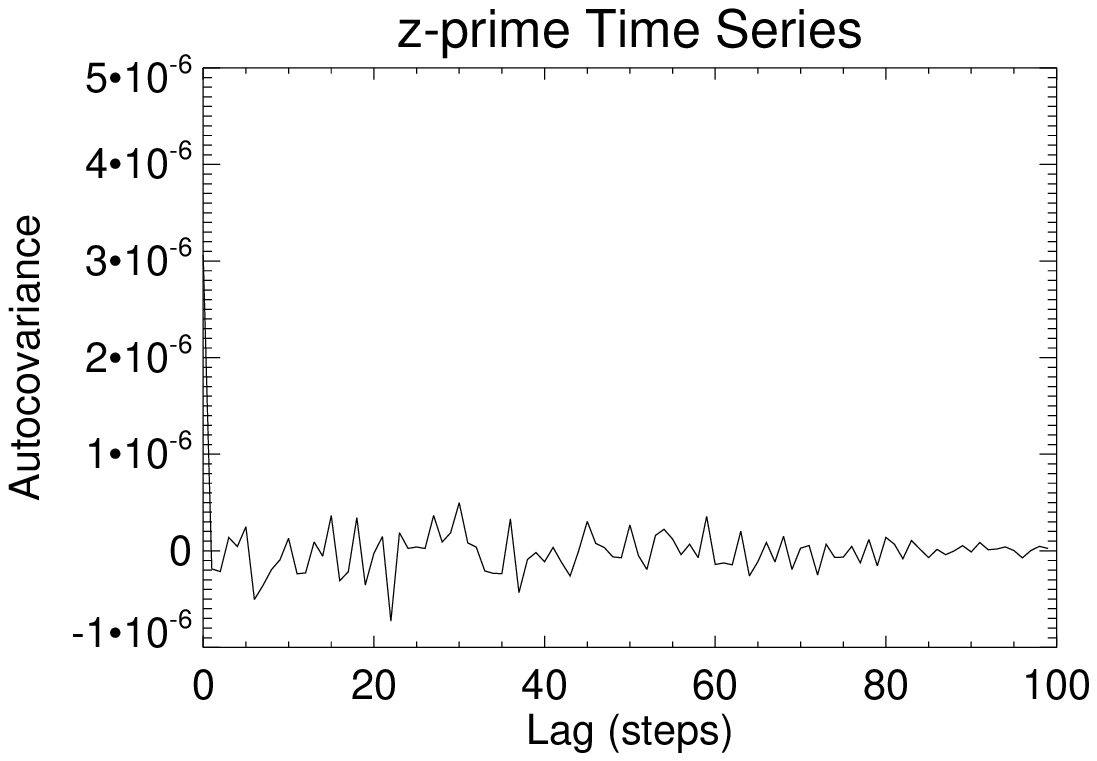}
\end{center}
\caption{{\it Left}: Autocovariance estimator of the residuals fit with a white noise model and a flat baseline (black  line) compared with the autocovariance estimator of a best-fit Gaussian Process model (orange line). These example light curves were for the SpeX 1.79 $\mu$m bin and SpeX 1.43 $\mu$m bin, which have stronger baseline trends, and the $z'$ filter for Jan 04, 2012 data, which was flatter. Appendix A shows simulated individual realizations of the same Gaussian Processes and how they compare to the covariance kernel. {\it Right}: The same light curves are fit with the Gaussian-Process model and the autocovariance of the final residuals (black line) show no correlation between data, just the white noise peak at zero lag.}\label{acfexamples}
\end{figure*}

The need for a covariance kernel is justified by the fact that the time series are not well fit by a flat baseline. If we do fit the time series to a flat, white noise baseline model -- with fixed semi-major axis, impact parameter and orbital period from literature values \citep{bean09} and free planet-to-star radius ratio $\rprs$ and free linear limb darkening -- the resulting residuals show correlations, as visible in the autocovariance estimator. If the autocovariance has a spike at zero lag and then is flat for all lags greater than zero, the noise is independent and identically distributed - white noise. On the other hand, if there is structure to the autocovariance, then there are correlation between flux measurements. Figure \ref{acfexamples} shows a few examples of the autocovariance estimator of the residuals and the autocovariance estimator of the best-fit Gaussian process model. The autocovariance estimator is a biased estimator \citep{wei06} so it can be different from the covariance kernel. Appendix A shows the kernel, individual realizations and the ensemble average of the autocovariance of the same best-fit hyperparameters used in Figure \ref{acfexamples}.

The inclusion of correlated noise requires that the full likelihood function must be used in evaluating a model instead of a plain $\chi^2$ statistic. The full likelihood function is
\begin{equation}
\mathcal{L} = \frac{1}{(2 \pi)^{n/2} |\mathbf{C}|^{1/2}}\mathrm{exp}\left(-\frac{1}{2}\mathbf{r}^T \mathbf{C}^{-1} \mathbf{r} \right)
\end{equation}
where $\mathcal{L}$ is the likelihood function when evaluating a model for covariance matrix $\mathbf{C}$ and residual vectors $r_n = y_n - f_n$ for data value $y_n$ and model value $f_n$ and $^T$ is the transpose \citep{gibsonGP}.
In the case of statistically independent non-correlated data $\Theta_0 = 0$ and $\mathcal{L} \propto \mathrm{exp}^{-\chi^2}$ where $\chi^2 = \sum\limits_n(y_n - f_n)^2/\sigma_n^2$, the standard chi-squared statistic. However, we find that $\Theta_0 \ne 0$ and that correlated noise is present in the data.

\subsection{Extracted Parameters}\label{paramExt}

We fit all time series with the transit function from \citet{magol} and use a series of MCMC chains to explore the parameter uncertainty distributions. The out-of-transit flux, planet-to-star radius ratio $\rprs$, linear limb darkening $u_1$ and hyper-parameters $\Theta_0$, $\Theta_1$ and $\Theta_2$ are fitted to the data while all other transit parameters -- impact parameter, semi-major axis and orbital period -- are fixed at the literature values from \citet{bean09}. For all parameters and hyper-parameters we use flat priors. All parameters and hyper-parameters are constrained by the likelihood function except in the case where the covariance strength hyper-parameter ($\Theta_0$) is much smaller than the white noise, $\sigma_n$. In these cases the time scale hyper-parameter ($\Theta_1$) is poorly constrained but does not strongly affect the $\rprs$ result over many orders of magnitude. For the continuous parameters, we use Gaussian proposal distributions from the current value and for the discreet kernel index hyper-parameter ($\Theta_2$), we use a uniform proposal distribution over the integers from 0 to 3.

Each time step in the MCMC chain requires a matrix inversion when evaluating the likelihood, which can make evaluation computationally expensive. To decrease chain evaluation time, the time series are binned to 100 time points for the nights of Dec 23, 2011 and Jan 04, 2012 with a resulting bin sizes of $\sim$3 minutes. Although this is comparable to OH variation timescales, increased number of bins did not give different results. For Dec 29, 2011 we use 50 time points to keep the timescales comparable. The chains are run to 6000 points each with the first 1000 points discarded to allow for convergence, comparable to \citet{gibson13}'s 5000 points with 1000 discards. The MCMC data series show stable parameter distributions beyond 1000 points. Three independent chains for all light curves are used to check for local minima and all final radius parameters and uncertainties agree between the chains to within 0.1\%.

\begin{figure*}[!ht]
\begin{center}
\includegraphics[width=0.8\textwidth]{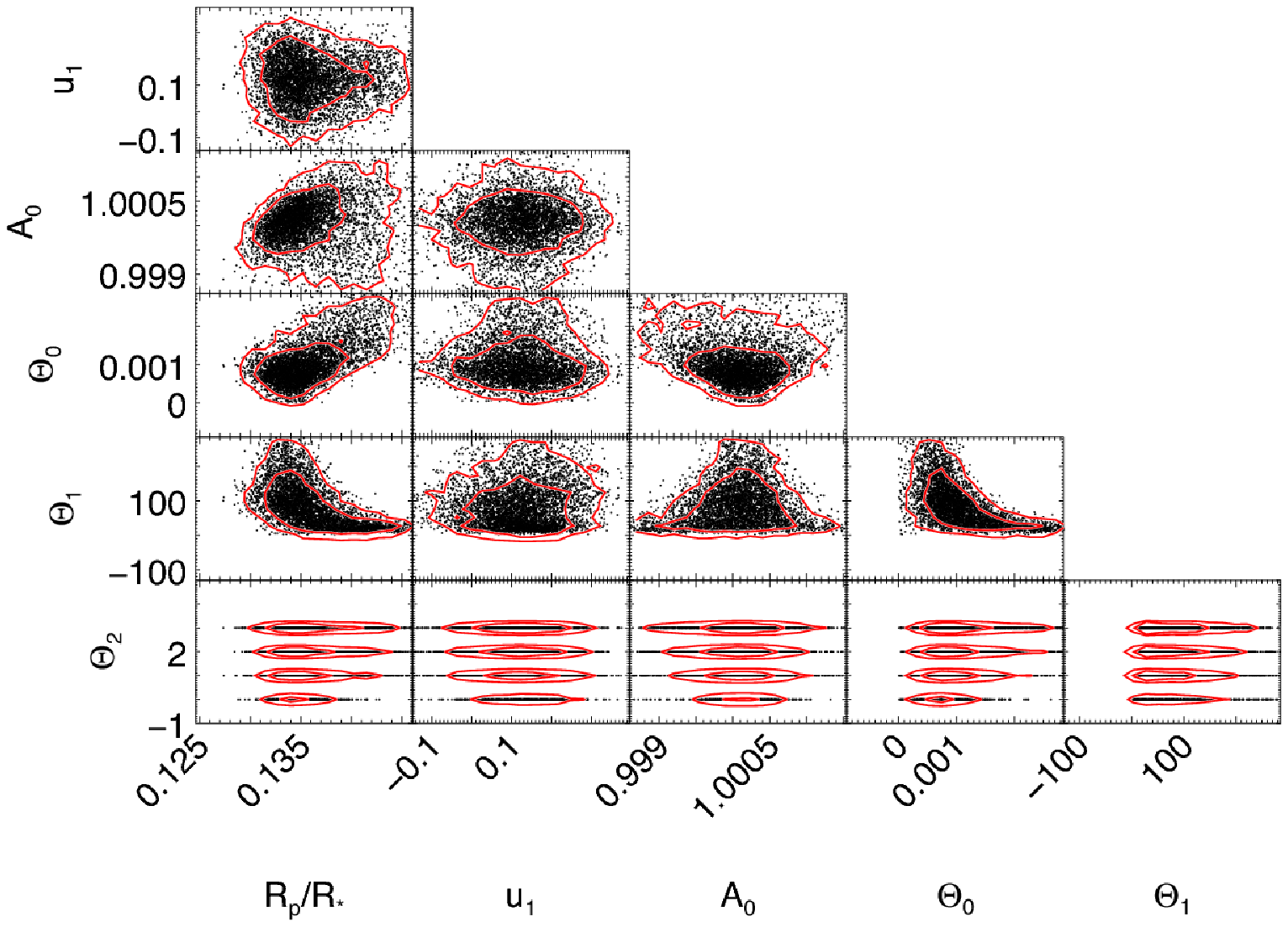}
\end{center}
\caption{Posterior density distribution for the fitted parameters for the night of Jan 04, 2012 for the MORIS $z'$ time series from the MCMC chain. The star to planet radius ratio $\rprs$ parameter correlates with the flux offset $A_0$, the hyper-parameters of the Gaussian process model $\Theta_0$ (strength of correlations) and $\Theta_1$ (inverse timescale of hyper-parameters) and $\Theta_2$ (the Mat\'{e}rn type) but not $u_1$ (the linear limb darkening parameter) because $\rprs$ and $u_1$ have nearly orthogonal distributions. The $\Theta_2$ parameter is a parametrization of the Mat\'{e}rn index $p$ and is discrete -- see Section \ref{gpmod} -- so there is an apparent discontinuity in phase space. 95\% and 68\% confidence regions for each projected distribution are shown in red. The correlation between parameters is smaller for the rest of the other SpeX and MORIS light curves.}\label{jan04cov}
\vspace{0.4in}
\end{figure*}

The example parameter correlation plot for the fit to the $z'$ light curve of Jan 04, 2012 in Figure \ref{jan04cov} shows how the fitted planet radius $\rprs$ can correlate to the other fitted parameters . This particular light curve showed the strongest dependence of $\rprs$ on the flux offset $A_0$ and hyper-parameters $\Theta_0$, $\Theta_1$ and $\Theta_2$. For the remaining curves, the $\rprs$ posterior is nearly orthogonal to the other hyper-parameters.

\begin{figure*}[!ht]
\begin{center}
\includegraphics[width=1.0\textwidth]{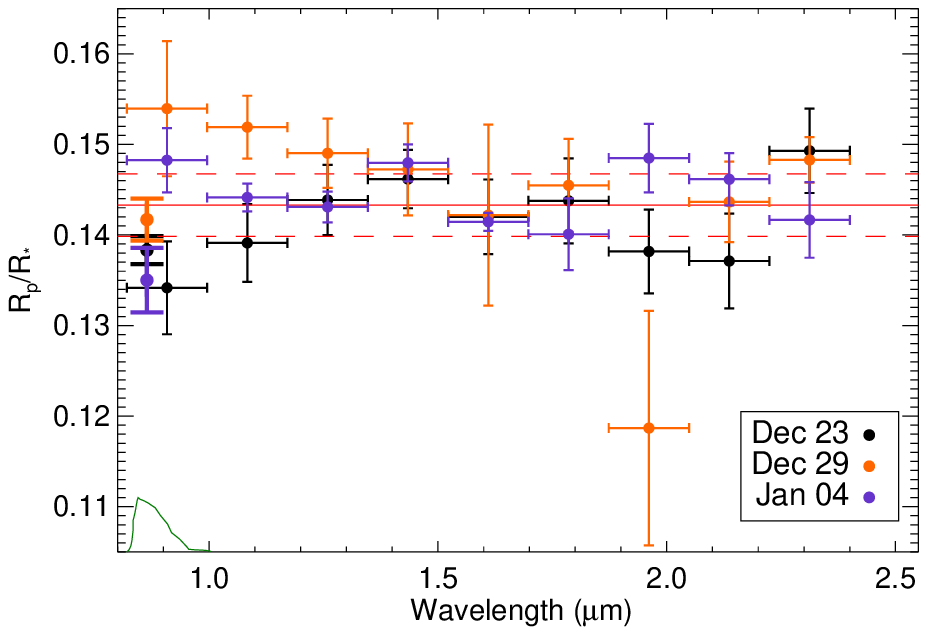}
\end{center}
\caption{Fitted radius ratio parameter as a function of wavelength for the three independent nights of observations with horizontal error bars as the bandwidths for spectral wavelength bins and vertical error bars with 68\% uncertainty. Points with bold lines are the simultaneous $z'$ photometry with the MORIS camera with a filter transmission curve (normalized to unity and scaled to 1/10 the plot size) shown in green. At a given wavelength, all points are within 2.1$\sigma$ of the weighted average, though there is a slight systematic shift downward for the night of Jan 04 (purple). The horizontal red line shows the CoRoT spacecraft radius \citep{bean09} and the dashed red lines indicate three scale heights above and below this value.}\label{3nightfig}
\vspace{0.5in}
\end{figure*}

The same \texttt{IRAF} data analysis pipeline and MCMC light curve fitting is applied to all three nights of observation and the fitted radius ratio and uncertainties are shown in Figure \ref{3nightfig}. The three nights are consistent within errors for a given wavelength. However, there is a slight decrease in radius fit for the night of Jan 04, 2012.

The three sets of observations in Figure \ref{3nightfig} are combined with a weighted average to produce a final transmission spectrum of the planet to be used in comparison to models. The weights are the inverse squared error in each wavelength bin for each night. We make the assumption that weather-related variability on the hot Jupiter itself has a negligible effect on the transmission spectrum. We also assume that the errors in radius from night to night are independent.

\begin{figure*}[!ht]
\begin{center}
\includegraphics[width=0.48\textwidth]{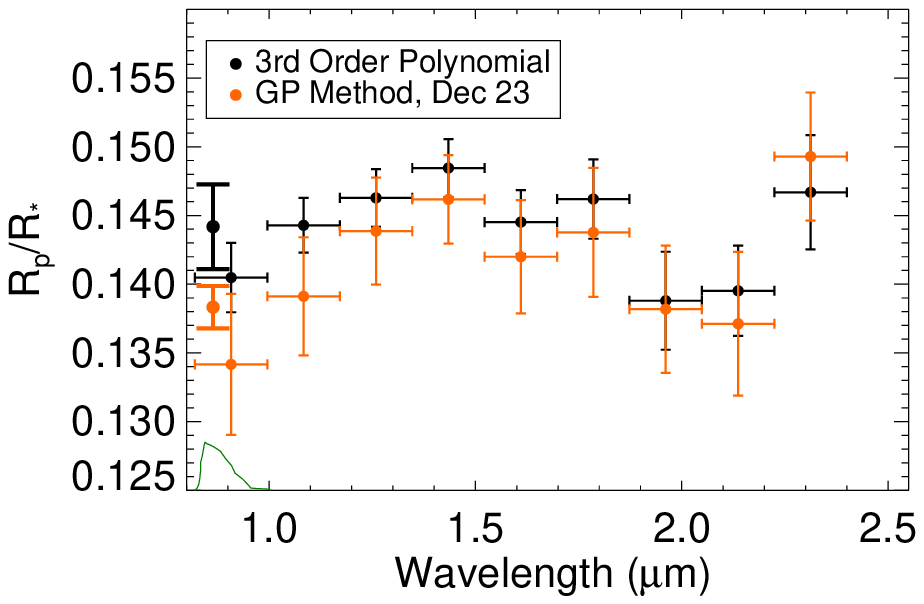}
\includegraphics[width=0.48\textwidth]{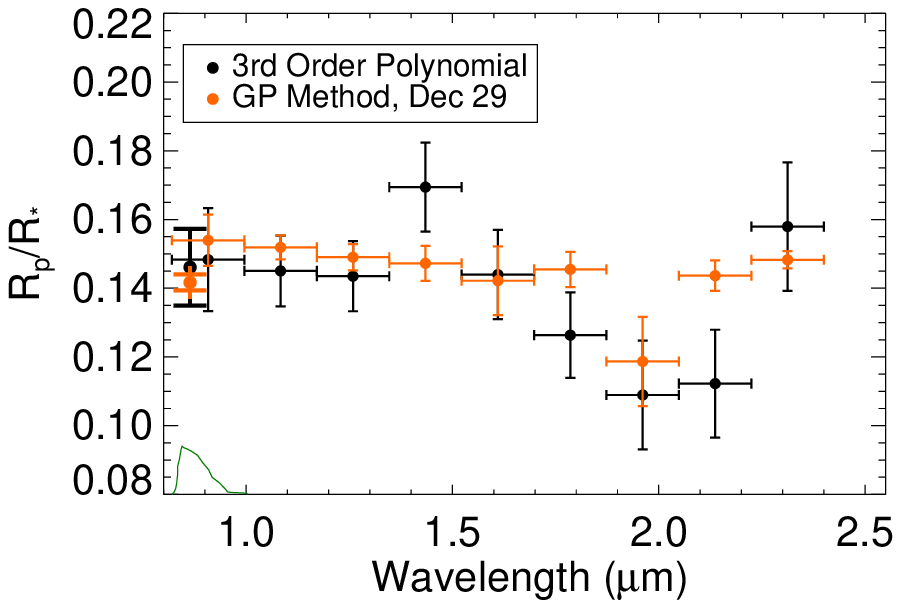}
\includegraphics[width=0.48\textwidth]{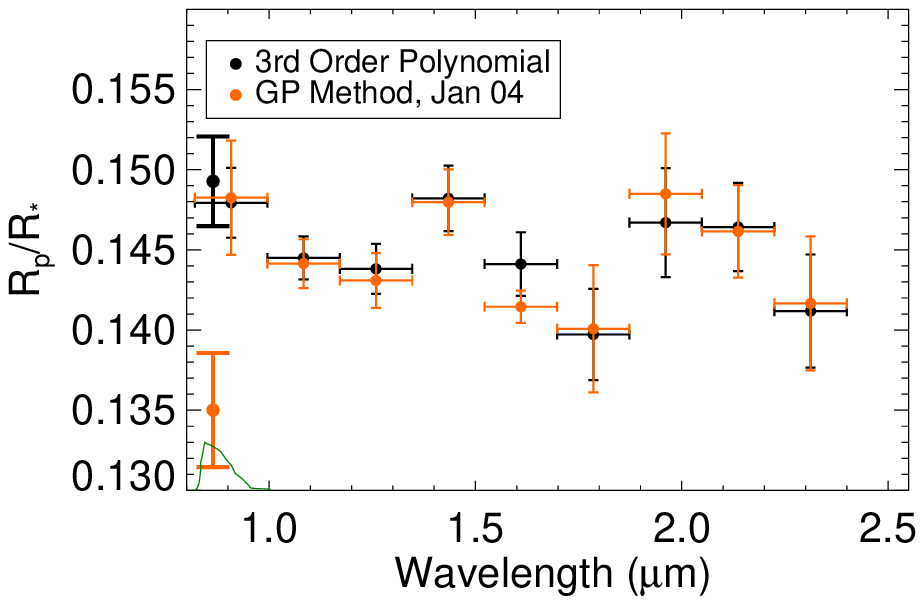}
\end{center}
\caption{Comparison between a polynomial baseline Levenberg-Marquardt fit and a  Gaussian process method for the baseline and flux variations. Photometry points ($z'$ band) are shown with bold lines, and the corresponding $z'$ bandpass is shown in green. The results are largely consistent for the SpeX data on the full transits of Dec 23, 2011 and Jan 04, 2012, but differ on the half transit of Dec 29, 2011 and the MORIS photometry for Jan 04, 2012. The MORIS photometry light curve for Jan 04, 2012 shows particularly large sensitivity to the fitting method because the flux bends down after egress -- see Figure \ref{jan04timeSer}. For the half transits of Dec 29, 2011, the third order polynomial (due to the shorter time baseline) produces much larger scatter for the half transit than the Guassian process method because it is fitting a specific shape to the light curve in the presence of red noise.}\label{polyVsGP}
\end{figure*}

It is worth noting that the Gaussian process method achieved higher precision than a polynomial baseline on the half-transit observation for Dec 29, 2011. Figure \ref{polyVsGP} shows a comparison between the best-fit radius when using a third order Legendre baseline fit as compared to the Gaussian correlated process. Both the scatter and error bars are larger when imposing a specific baseline shape. There was one particular light curve, the MORIS $z'$ photometry for Jan 04, 2012, that showed a very strong dependence on the type of treatment of systematic errors. As seen in the time series, Figure \ref{jan04timeSer}, the flux bends downward after egress. When the light curve is fit with a third order Legendre polynomial, this drop in flux is extrapolated to a higher flux during transit and thus the planet-to-star radius ratio estimate $\rprs$ is large. When the light curve is fit with the Gaussian Process method, the deviations from a flat baseline are best-fit with shorter time scale correlations and an essentially flat baseline. Our Gaussian process kernel (Equation \ref{eq:GPkern}) incorporates different shapes through the Mat\'{e}rn index, but does not increase the upper limit to the same value as the polynomial baseline. We adopt the Gaussian Process model fits, but given the dependence of $\rprs$ on the method, we also evaluate our science results with the polynomial model fits.

The average spectrum, listed in Table \ref{radTab}, has $\rprs$ uncertainties ranging from 0.7\% to 2\% of the mean value ($R_\mathrm{p,mean}/R_*$ = 0.144) across the near-infrared coverage. This uncertainty is comparable to the scale height of the atmosphere ($\sim$0.8\% for a 2400K atmosphere), whereas strong spectral features are expected to be three to five scale heights in planet radius variation \citep{burrowsChapter}. Figure \ref{3nightfig} shows no immediate statistically significant (5$\sigma$) molecular features.

\begin{table}
	\begin{center}
	\begin{tabular}{c*{1}{c}}
	Wavelength     & $\rprs$	  \\
	($\mu$m)	       &					 \\
	\hline \hline
	   z' (0.86)	       &	 0.1389	$\pm$	0.0012$^a$	\\
	   0.908 	       &	 0.1450	$\pm$	0.0027		\\	   
	   1.083		&	 0.1448	$\pm$	0.0013		\\
	   1.259 		&	 0.1440	$\pm$	0.0014		\\
	   1.434		&	 0.1474	$\pm$	0.0016		\\
	   1.610		&	 0.1415	$\pm$	0.0010		\\
	   1.786		&	 0.1426	$\pm$	0.0026		\\
	   1.961		&	 0.1431	$\pm$	0.0029		\\
	   2.137		&	 0.1440	$\pm$	0.0022		\\
	   2.312		&	 0.1470	$\pm$	0.0020		\\
	\end{tabular}
	\end{center}
	\caption{\rm  Weighted average planet-to-star radius ratio $\rprs$ for the three nights of observations shown in Figure \ref{3nightfig}. Quoted error bars are calculated by propagating the individual MCMC uncertainties in quadrature. The central wavelength for each 0.1755 $\mu$m bin is given in the first column except for the photometry filter where the first moment is given in parentheses. $^a$ the MORIS $z'$ time series showed particularly large sensitivity to the treatment of systematic errors. A polynomial baseline fit gives a weighted average $\rprs = 0.147 \pm 0.002$}\label{radTab} \raggedright

\end{table}

\section{Comparison with Models}\label{tdepths}

The error-weighted average transmission spectrum for the three nights is compared against a representative model for hot Jupiter atmospheres from \citet{fortney08,fortney10}. We select this model as a starting point because it has a published infrared spectrum, solar abundances and equilibrium chemistry. The blackbody temperatures fit to infrared data of $\approx$2400 K \citep[$T_{bb}$=2380 K,$T_{bb}$=2460K][]{zhao12,deming11}, and short orbital period $P$ = 1.509 days \citep{barge08} indicate that it is comparable to the $T_\mathrm{kinetic}$=2500 K isothermal model from \citet{fortney10}.

\begin{figure*}[!ht]
\begin{center}
\includegraphics[width=\textwidth]{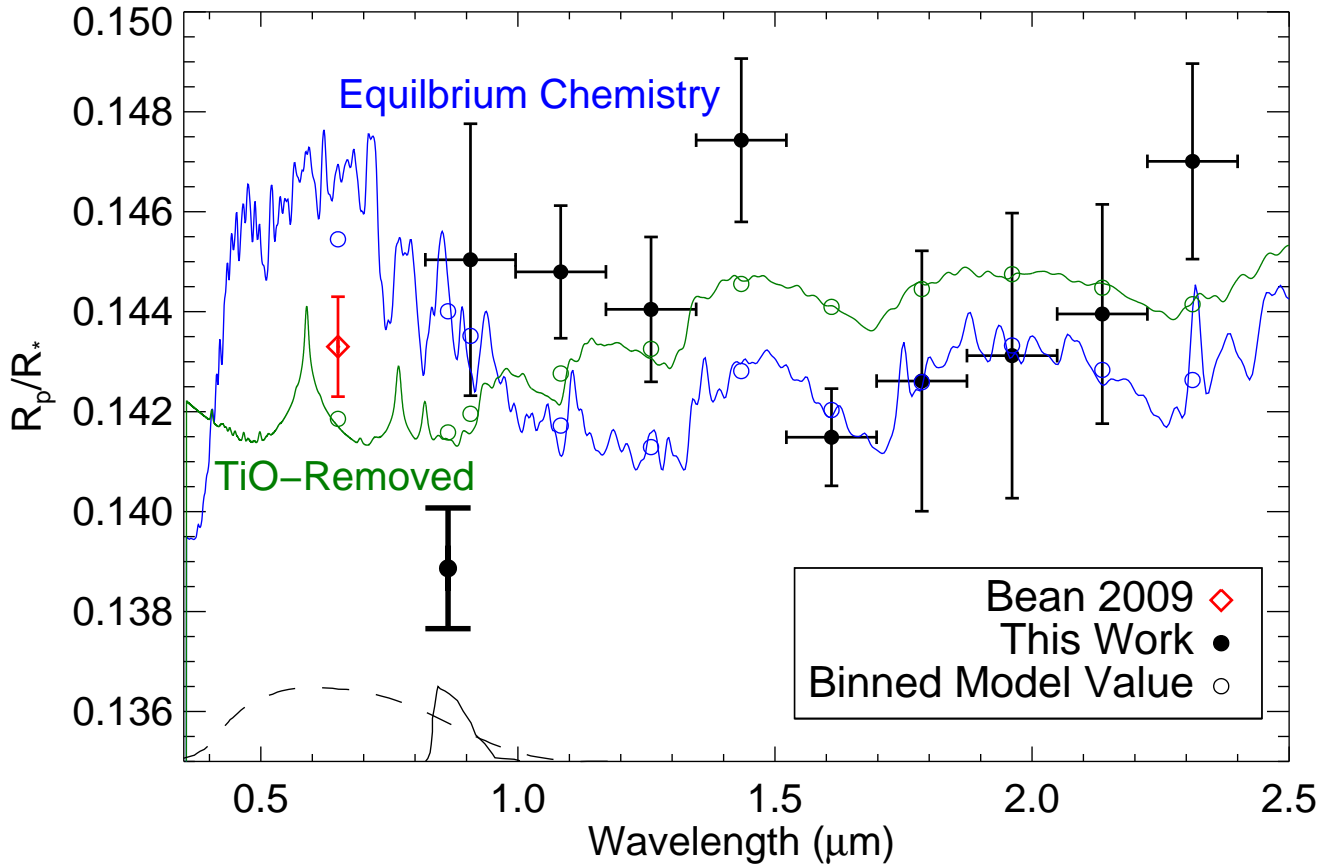}
\end{center}
\caption{Measured planet-to-star radius ratio spectrum compared to a 2500 K isothermal model (blue) from \citep{fortney10} with no clouds or hazes but significant TiO/VO absorption. Black data points are the weighted average IRTF data for three transits with spectral data in thin lines and photometry data in thick lines. The red point is the CoRoT value from 36 transits \citep{bean09}. Y error bars represent 1$\sigma$ uncertainties whereas X error bars span spectral windows, except in the cases of photometric data. Photometric filter curves that are normalized to unity and scaled to 1/10 of the figure are shown in with black lines for the CoRoT planet finder response (dashed line) and the MORIS $z'$ filter (solid line). The IRTF data (black) combined with the CoRoT point (red) disfavor the TiO/VO-driven optical to infrared absorption slope and give a $\chi^2$ per degree of freedom of 4.6 as compared to a $\chi^2$ of 2.4 per degree of freedom for the same model with TiO removed (green).}\label{modelTn}
\end{figure*}

The equilibrium model from \citet{fortney08,fortney10} shows substantial opacity in the optical as compared to the infrared due to mainly TiO and VO absorption, so we compare the CoRoT derived radius \citep{bean09} to our transmission spectrum, as seen in Figure \ref{modelTn}. The \citet{bean09} radius is larger than the original discovery \citep{barge08}, but we adopt the \citet{bean09} value because it was found with a newer data processing pipeline.  The combined CoRoT data and IRTF data show no evidence for an optical to infrared slope. Fitting a flat spectrum to the data gives a reduced chi-squared ($\bar{\chi^2}$) of 2.9 for 10 degrees of freedom whereas the model with TiO/VO gives $\bar{\chi^2}$ of 4.6 for 10 degrees of freedom. The same model with TiO and VO artificially removed, gives $\bar{\chi^2}$ of 2.4 for 10 degrees of freedom. As mentioned in Section \ref{paramExt}, the MORIS results were particularly sensitive to the choice of model to fit the time series. If we use a polynomial baseline fit to the time series, the TiO/VO rich model is again disfavored with a $\bar{\chi^2}$ of 2.9 as compared to the TiO-removed model with $\bar{\chi^2}$ of 1.6 and a flat line of $\bar{\chi^2}$ of 1.3.

The hot Jupiter WASP-19b also shows no evidence for TiO/VO absorption \citep{huitson13,mancini13}. For this planet, TiO/VO depletion is expected since WASP-19b has no observed temperature inversion (stratosphere) \citep{anderson13}. WASP-12b similarly has no stratosphere, but does have a larger optical to infrared transit depth ratio. Models for WASP-12b that included either TiO/VO or TiH were consistent with initial data \citep{swain2013,stevenson2013wasp12} but adding optical data and including models with aerosols together suggest that WASP-12b has low levels of TiO/VO \citep{sing2013}.

CoRoT-1b, by contrast, is better matched by models {\it with} a stratosphere or isothermal temperature profile. \citet{rogers09} compare a suite of equilibrium abundance models with multi-color photometric secondary eclipses on CoRoT-1b. The molecular features in these models appear in absorption or emission depending on the temperature structure of a planet's atmosphere and \citet{rogers09}'s models with no temperature inversion fail to produce the $Ks$ and narrowband 2.1 $\mu$m brightness temperatures for the planet. The only models that come close to matching the observations include an extra optical absorber at the 0.01 to 0.1 bar level. \citet{deming11} also find that the secondary eclipse fluxes are better fit with models that include a temperature inversion than models without. Still, \citet{deming11} find consistency with a blackbody spectrum, which could be due to an isothermal profile or a thick layer of high altitude dust.

Plausible absorbers that could create a stratosphere in CoRoT-1b are TiO and VO \citep{fortney08twoclass}, which should also increase the optical radius as compared to the infrared. However, since our IRTF-CoRoT combined spectrum is disfavored by models with TiO/VO absorption, we expect another species is responsible for the temperature inversion, such as sulfur-containing compounds \citep{zahnle09}. Alternatively, a high altitude haze or dust \citep[e.g.,][]{pont13} could explain the blackbody-like emission from CoRoT-1b and also flatten out molecular features in the transmission spectrum. 

Many other atmospheric optical scattering and absorbing processes may occur in hot Jupiter atmospheres including (a list from \citet{sing2013}): Raleigh scattering off molecules, Mie and Raleigh scattering off dust, tholin hazes and gray absorbing clouds. The majority of these processes increase planetary radii at short wavelengths as compared to long wavelengths. Our observations, by contrast, show that the optical radius is not significantly larger than the infrared radius based on the CoRoT photometry. Gray absorbing clouds are the one item on the above list that could equalize the optical and infrared transit depths. Recent observations of HAT-P-32b \citep{gibson13clouds}, HAT-P-12b \citep{line13hatp12}, and Kepler-7b \citep{demoryClouds} indicate that high altitude clouds may be pervasive in exoplanet atmospheres. In HAT-P-32b, gray-absorbing clouds may obscure TiO/VO features (or the TiO and VO may be present at very low abundances) \citep{gibson13clouds}. Analysis of the above a processes is limited with only CoRoT photometry, but additional optical spectroscopy would be useful in constraining the strength of these scattering and absorbing phenomena.

We observe a 2$\sigma$ peak at 1.4 $\mu$m in the spectrum, close to a 1.4 $\mu$m water feature seen in all temperature classes of \citet{fortney10}'s equilibrium models. This same feature was used to detect water vapor in WASP-19b with HST \citep{huitson13}. However, if water vapor in CoRoT-1b caused the 2$\sigma$ feature at 1.4 $\mu$m, the 1.8 $\mu$m radius should also be elevated, which is not seen in our spectrum. One possible explanation is that the 1.4 $\mu$m peak is due to H$_2$C$_2$ or HCN, which are both predicted to be abundant in hot Jupiter atmospheres \citep{moses13}. Unfortunately, the significance level of this peak is too low to distinguish between these molecules or rule out the possibility of a statistical deviation or un-removed telluric absorption signature.

\section{Conclusion}

We present a 0.8 $\mu$m to 2.4 $\mu$m transmission spectrum for the hot Jupiter CoRoT-1b, the faintest ($K$=12.2) host star for which the planet has been spectroscopically characterized to date. With the MOS method and a single nearby simultaneous reference star, we achieve 0.03\% to 0.09\% precision of the transit depth $R_\mathrm{p}^2/R_*^2$ when combining all three nights of data, comparable to one atmospheric scale height for this hot Jupiter's temperature. We conclude the following items from our analysis:

\begin{itemize}

	\item The IRTF spectrum, when combined with the optical planet-to-star radius ratio derived from observations by the CoRoT spacecraft \citep{bean09}, disfavors a model that includes TiO/VO as compared to a model that is spectrally flat or has TiO removed. This goes against the prediction that CoRoT-1b's thermal inversion is due to TiO/VO absorption. Other recently characterized hot Jupiters with similarly high temperatures, WASP-19b and WASP-12b, also lack strong TiO/VO \citep{anderson13,sing2013} features, but TiO/VO is expected to be depleted in these planets because they have no observed temperature inversions.

	\item No statistically significant molecular features are seen in the 0.8 $\mu$m to 2.4$\mu$m transmission spectrum, although there is a small 2$\sigma$ peak at 1.4 $\mu$m, possibly due to H$_2$C$_2$ or HCN. Our precision is not high enough to constrain the detailed composition of H$_2$O, CO, and other gases due to the systematics and faintness of the host star.

	\item The Gaussian process method for determining systematics and the baseline achieves better precision in extracted parameters and more robustness when applied to the half-transit on Dec 29, 2011 as compared to a deterministic polynomial baseline. For data sets with strong out-of-transit curvature, the Gaussian Process model can give significantly different results from a polynomial baseline.
	
\end{itemize}

\bibliography{ms}

\section{Acknowledgements}
This research was funded in part by the New York/NASA Space Grant Fellowship. Based on observations made from the Infrared Telescope Facility, which is operated by the University of Hawaii under Cooperative Agreement no. NNX-08AE38A with the National Aeronautics and Space Administration, Science Mission Directorate, Planetary Astronomy Program. M.Z. is supported by the Center for Exoplanets and Habitable Worlds at the Pennsylvania State University. Thanks to Eva-Maria Mueller and Joyce Byun for useful MCMC suggestions. The authors wish to recognize and acknowledge the very significant cultural role and reverence that the summit of Mauna Kea has always had within the indigenous Hawaiian community.  We are most fortunate to have the opportunity to conduct observations from this mountain. We also thank the anonymous referee for useful comments and corrections.

\clearpage
\section{Appendix A: Simulated Series}

\begin{figure*}[!ht]
\begin{center}
\includegraphics[width=0.5 \textwidth]{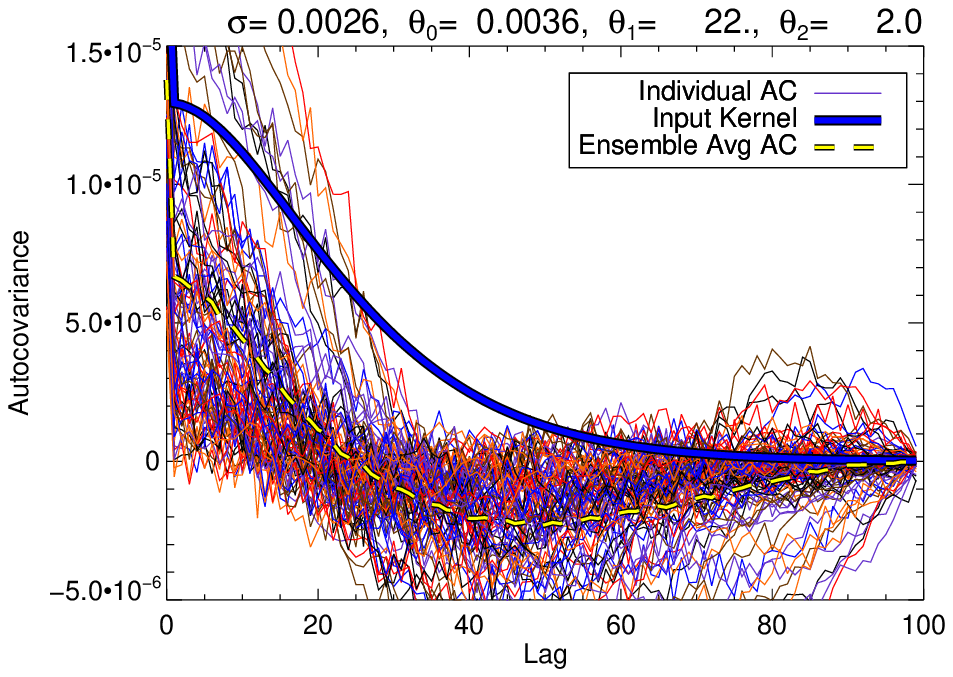}
\includegraphics[width=0.5 \textwidth]{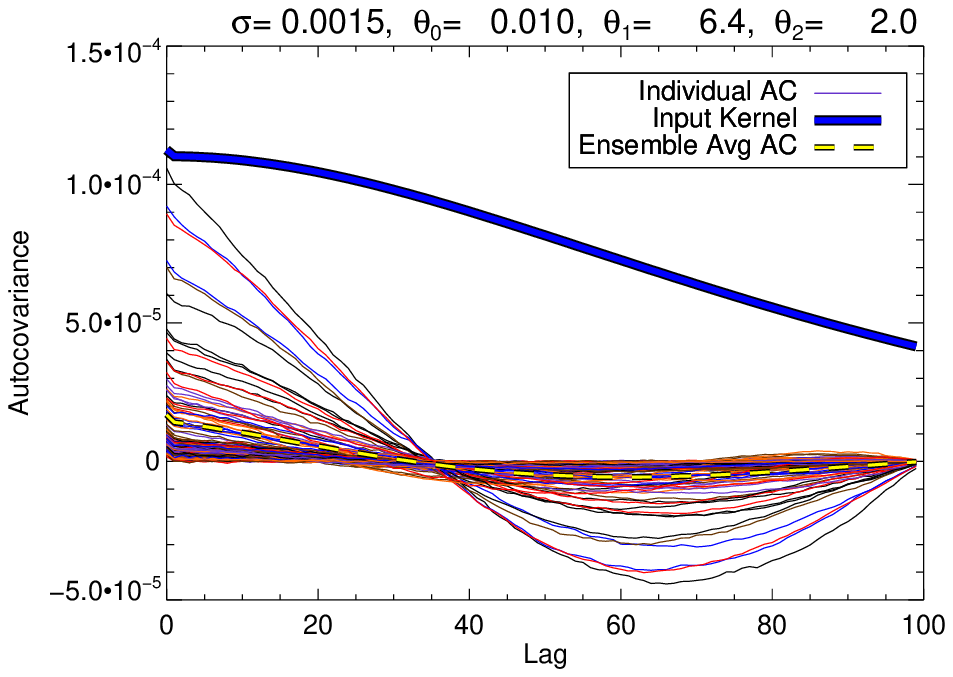}
\includegraphics[width=0.5 \textwidth]{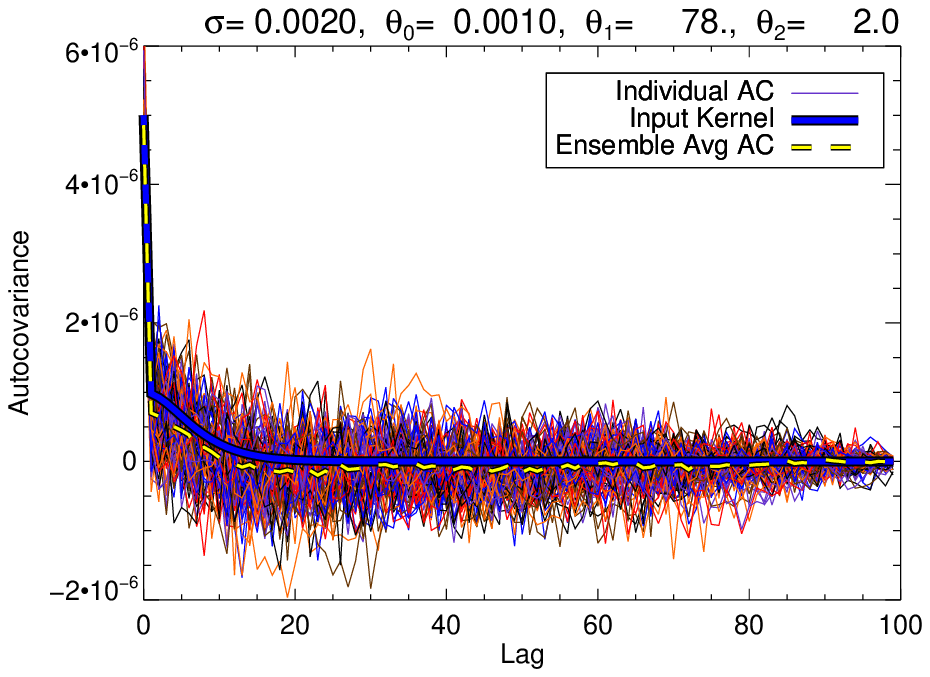}
\end{center}
\caption{Autocovariance estimators of simulated time series for a given kernel (solid blue line) and hyper-parameters given in the title. The individual colored thin lines show different realizations of the same covariance kernel and the average is shown as a dotted yellow line. The ensemble average is not equal to the kernel function because the autocovariance estimator is a biased estimator of the true autocovariance \citep{wei06}. }\label{simAutoC}
\end{figure*}

In order to compare an autocovariance of residuals to an input kernel, it is illustrative to show the autocovariance of some simulated time series. Figure \ref{simAutoC} shows simulations for the best-fit hyper-parameters from the 1.79 $\mu$m, 1.43 $\mu$m and $z'$ light curves on the night of January 04, 2012. The two autocovariance plots of the residuals are shown in Figure \ref{acfexamples}

\end{document}